\documentclass[aps,prb,preprint,superscriptaddress,amsmath,amssymb]
{revtex4-1}
\usepackage{graphicx}
\usepackage{dcolumn}
\usepackage{bm}

\newcommand{\ks} {{\bf k}}
\newcommand{\vs} {{\bf v}}

\newcommand{\ps} {{\bf p}}
\newcommand{\ds} {{\bf d}}
\newcommand{\pc} {{\bf P}}
\newcommand{\rs} {{\bf r}}

\newcommand{\es} {{\bf e}}
\newcommand{\ac} {{\bf A}}
\newcommand{\ec} {{\bf E}}
\newcommand{\gc} {{\bf G}}
\newcommand{\jc} {{\bf J}}
\newcommand{\eps} {{\varepsilon}}

\begin{document}


\title{Ultrafast energy absorption and photoexcitation of bulk plasmon
in crystalline silicon subjected to intense near-infrared ultrashort laser pulses}

\author{Tzveta Apostolova}

\address{Institute for Nuclear Research and Nuclear Energy, Bulgarian Academy of Sciences, Tsarigradsko chausse 72, 1784 Sofia, Bulgaria}
\address{Institute for Advanced Physical Studies,
New Bulgarian University, 1618 Sofia, Bulgaria}

\author{Boyan Obreshkov}

\address{Institute for Nuclear Research and Nuclear Energy, Bulgarian Academy of Sciences, Tsarigradsko chausse 72, 1784 Sofia, Bulgaria}

\author{Iaroslav Gnilitskyi}

\address{NoviNano Lab LLC, Pasternaka 5, 79015 Lviv, Ukraine}
\address{Department of Photonics, Lviv Polytechnic National University,Stepana Bandery 14, 79000, Lviv, Ukraine}
\address{University of Modena and Reggio Emilia (UNIMORE), Amendola 2, 42122 Reggio Emilia, Italy}

\begin{abstract}

We investigate the non-linear response and energy absorption in bulk silicon irradiated by intense 12-fs near-infrared laser pulses. Depending on the laser intensity, we distinguish two regimes of non-linear absorption of the laser energy: for low intensities, energy deposition and photoionization involve perturbative three-photon transition through the direct bandgap of silicon. For laser intensities near and above 10$^{14}$ W/cm$^2$, corresponding to photocarrier density of order 10$^{22}$ cm$^{-3}$, we find that absorption at near-infrared wavelengths is greatly enhanced due to excitation of bulk plasmon resonance. In this regime, the energy  transfer to electrons exceeds a few times the thermal melting threshold of Si. The optical reflectivity of the photoexcited solid is found in good qualitative agreement with existing experimental data.
In particular, the model predicts that the main features of the reflectivity curve of photoexcited Si as a function of the laser fluence are determined by the competition between state and band filling associated with Pauli exclusion principle and Drude free-carrier response. The non-linear response of the photoexcited solid is also investigated for irradiation of silicon with a sequence of two strong and temporary non-overlapping pulses. The cumulative effect of the two pulses is non-additive in terms of deposited energy. Photoionization and energy absorption on the leading edge of the second pulse is greatly enhanced due to free carrier absorption.

\end{abstract}

\maketitle

\section{Introduction}
\label{S:1}

Time-resolved optical experiments on femtosecond laser excited dielectrics \cite{Shank,Tom,Sokolowski-Tinten,Glezer,Siegal,Huang,Rousse,Kim,Sokolowski, Bonse,Harb} provide evidence that ultrafast solid to liquid phase transition occurs after a large amount of laser energy is deposited in the solid material during a time interval much shorter than the thermalization of the absorbed energy. The photoexcitation of a critical number of electron- hole pairs results in bond softening and structural phase transition. Theoretical models \cite{Sarma,Stampfli,Silvestrelli,Darkins} were developed aiming to investigate electronically-driven ultrafast melting mechanism in semiconductors. These studies predict that lattice instability in the dense plasma develops once the critical density of electron-hole pairs is of order 10$^{22}$ cm$^{-3}$. For instance in silicon irradiated with visible wavelengths, such high densities are reached at the fluence of about 0.2 J/cm$^2$ \cite{Sokol}.
 
The high density plasma of photoexcited charge carriers leads to distinct change of the optical reflectivity. Changes of optical constants of the strongly excited dielectrics are measured using pump-probe spectroscopy techniques \cite{Glezer,Sokol,Hulin,Frigerio}. The optical reflectivity of photoexcited silicon as a function of the pump pulse fluence was measured in \cite{Sokol} for the 625 nm wavelength. At very early times (150 fs after the excitation), when the plasma is not yet thermalized, the experimental data showed that the reflectivity initially decreases for relatively low pump fluences until it reaches a minimum at 0.2 J/cm$^2$. By increasing the pump fluence above that value, a sharp rise of reflectivity of the probe pulse was observed. The fluence dependence of the reflectivity curve was consistently interpreted within a simplified Drude model for the dielectric function of the optically excited Si, i.e. assuming that the optical properties are dominated by free-carrier response.

Dielectric properties of photoexcited solids are often modeled with Drude response of free carriers embedded in a dielectric medium\cite{Rethfeld,Medvedev}. More elaborate model for the macroscopic dielectric function of the laser-excited state incorporating Pauli blocking and screening of the Columbic electron-hole attraction was proposed in \cite{Benedict,Spataru}. Good qualitative agreement with measured dielectric function of gallium arsenide was found. More recently, first principle approaches based on time-dependent density functional theory (TDDFT) have been developed and applied for understanding the optical properties of strongly excited semiconductors and dielectrics. Ultrafast optical breakdown thresholds of dielectrics such as diamond and silica were obtained using TDDFT \cite{Otobe,OtobeT}. With the increase of the light pulse intensity, the number of photoexcited charge carriers increases and dense plasma of electron-hole pairs is established. When the corresponding plasma frequency matches the laser frequency, a resonant energy transfer occurs from the light pulse to the electrons and dielectric breakdown occurs associated with sharp rise of the optical reflectivity.

In \cite{Yabana}, the reflectivity of photoexcited silicon was calculated as a function of the peak laser intensity and a qualitative agreement with the experimental observation in \cite{Sokol} was found. Consistent interpretation of the intensity dependence of the reflectivity curve was given in terms of simplified free electron Drude model. In \cite{Sato,Sato2}, the dielectric response of the crystalline silicon following irradiation by a high intensity near-infrared laser pulse was obtained from numerical pump-probe experiments. The results showed that the optical response of the photo-excited silicon exhibits characteristic features of electron-hole plasma in non-equilibrium phase. The real part of the dielectric function was found to be well described by a Drude free-carrier response with screened plasma frequency determined from ground state properties. The effective mass of the charge carriers was found to increase monotonically with the increase of the laser intensity. Optical anisotropy in the response of the photoexcited solid was also reported. 

While standard pump-probe spectroscopy studies electronic dynamics with femtosecond time resolution, advances in laser technologies resulted in the attosecond metrology \cite{Krausz,Stockman}. In nonlinear attosecond polarization spectroscopy \cite{Sommer}, the oscillating laser electric field is used to measure the non-linear polarization which in turn determines the amount of energy reversibly or irreversibly exchanged between the electromagnetic field and the dielectric material. Nonlinear polarization spectroscopy yields more complete information about the dynamic electronic response to strong fields with attosecond time resolution. For instance, measurements utilizing attosecond spectroscopy in combination with TDDFT calculations allowed to resolve electron dynamics in crystalline silicon \cite{Schultze}. Interest in the time evolution of the excitation process is motivated by applications like petahertz signal processing \cite{Krausz2,Han} and mechanisms of ultrafast dielectric breakdown \cite{Du,Stuart,Kautek,Lenzner,Li,Apostolova, Apostolova2,Rajeev,Zhokhov}.

The optical breakdown thresholds are usually associated with femtosecond laser ablation \cite{Geohegan:2010qr,Gamaly,Gamaly2,Mirza}. During femtosecond laser ablation laser-induced periodic surface structures (LIPSS) form under certain irradiation conditions. Typically LIPSS occur after irradiation by multiple laser pulses. Two of the main mechanisms of LIPSS formation are attributed to interference between laser induced surface-plasmon-polariton waves with the incident laser  \cite{Bonse1} or to the development of hydrodynamic instability in the molten surface layer \cite{Tsibidis,Gurevich}. The plasmon mechanism of ripple formation is due to the spatial modulation of the deposited energy in the surface layer as a result of interference between the incident and a surface-scattered wave \cite{Ashkenasi} intermediated by surface roughness.  A grating assisted interference between the driving laser and photoinduced surface plasma oscillations was demonstrated in optical reflectivity measurements of photoexcited silicon surface \cite{Miyaji}. 

Advances in technologies found a regime where highly regular LIPSS are produced by a few temporally overlapping identical femtosecond laser pulses with above threshold fluence \cite{Gnilitskyi,Shugaev} in contrast to the standard processing procedure utilizing multiple laser pulses with near threshold fluences. Such approach allows to substantially improve production rate of nanostructures on large surface area.

In this paper we report the time evolution of the conduction electron density and excitation of collective plasma oscillations of the electron gas in crystalline silicon irradiated by intense ultrashort near infrared laser pulse. We distinguish two regimes of non-linear absorption: low-intensity regime when laser energy is transferred to electrons by three- and four-photon transitions, and high-intensity regime when free-carrier absorption becomes prominent. Scaling laws of the absorbed energy and electron density as a function of the laser intensity are obtained. Optical constants (refractive index, exctinction coefficient)and the normal incidence reflectivity of the photo excited silicon are obtained and compared to existing experimental data and other theoretical works.
Photoexcitation by a double pulse sequence resulting in highly efficient energy transfer is also demonstrated. Unless otherwise stated, atomic units are used throughout this paper ($e=\hbar=m_e=1$).

\section{Theoretical formalism}
\label{S:2}

In velocity-gauge, the Schr\"{o}dinger equation in single-active electron approximation is

\begin{equation}
  i\partial_t|\psi_{v\ks} (t)\rangle= H(t)|\psi_{v\ks}(t)\rangle
\end{equation}
here $v\ks$ labels the initially occupied Bloch states in the valence band with definite crystal momentum $\ks$, $H(t)$ is the time-dependent Hamiltonian
\begin{equation}
  H(t)=\frac{1}{2} [\ps+\ac(t)]^2+V(\rs),
\end{equation}
In the empirical pseudopotential method \cite{Cohen}, the periodic ionic lattice potential is presented by a plane wave-expansion in the basis of reciprocal lattice wave-vectors
\begin{equation}
 V(\rs)= \sum_{\gc} V(G) \cos(\gc \cdot \boldsymbol \tau) e^{i \gc \cdot \rs}
\end{equation}
where 2 $ \boldsymbol \tau=a_0/4(1,1,1)$ is the relative vector connecting two Si atoms in a crystal unit cell and $a_0=$ 5.43 \AA ~is the bulk lattice constant. The pseudopotential formfactors (in Rydberg) are $V(G^2=3)=$-0.21, $V(G^2=8)=$0.04 and $V(G^2=11)=$0.08.

The macroscopic vector potential is split into applied laser and induced vector potentials $\ac(t)=\ac_{{\rm ext}}(t)+\ac_{{\rm ind}}(t)$, the total pulsed electric field is $\ec(t) =-d \ac/dt$.  The applied vector potential is related to the electric field of the incident laser  by $\ec_{{\rm ext}}=-d \ac_{{\rm ext}}(t)/dt$, which we parameterize by a temporary Gaussian function
\begin{equation}
\ec_{{\rm ext}}(t)=\ec e^{-\ln(4) (t-t_0)^2/\tau_L^2} \cos \omega_L t
\end{equation}
where $\ec$ is a unit vector in the direction of
the laser polarization, $\omega_L$ is the laser frequency
corresponding to photon energy $\hbar \omega_L$,
$\tau_L$ is the pulse length and $t_0$ is the position of the pulse peak. The vector potential $\ac_{{\rm ind}}(t)$ is a result of the induced dipole moment per unit volume $\pc(t)$, which we determine self-consistently by solving the Maxwell's equation in long wavelength approximation
\begin{equation}
 \frac{d^2 \ac_{{\rm ind}}(t)}{dt^2}=4 \pi \jc(t)  ,
\end{equation}
here $\jc(t)$ is the macroscopic electric current density
\begin{equation}
\jc(t)=\sum_v \int_{{\rm BZ}} \frac{d^3 \ks}{4 \pi^3}
\langle \psi_{v\ks}(t) | \vs(t) |
\psi_{v\ks}(t) \rangle
\end{equation}
and $\vs(t)=\ps+\ac(t)$ is the velocity operator. The instantaneous polarization is $\pc(t)=\int^t dt' \jc(t')$, the absorbed laser energy per unit volume at time $t$  is calculated from the work done by the pulsed laser field in moving the electrons
\begin{equation}
 \Delta E(t) = \int_{-\infty}^t dt' \ec(t') \cdot \jc(t')
\end{equation}

\subsection{A.	Linear response of photoexcited Si}

In the remote past $t \rightarrow -\infty$, the laser vector potential
vanishes $\ac_{{\rm ext}}= \bf 0$ and the Maxwell's equation exhibits
a trivial solution with $\ac=\bf 0$. The electrons are in a ground state characterized by occupation numbers of Bloch states $f^0_{n \ks}=\{0,1\}$. However in the remote future $t \rightarrow + \infty$, when the applied laser vector potential has vanished $\ac_{{\rm ext}}= \bf 0$, the Maxwell's
equation may exhibit a non-trivial solution with $\ac(t) \ne \bf 0$. In this case, electrons are driven by the self-induced polarization.
The single-particle density matrix evolves in time according to
\begin{equation}
 i \frac{d \rho}{dt} = [H,\rho].
\end{equation}
where $H=H_0+V$, here $H_0$ is the field-free Hamiltonian and
$V= \ac\cdot \ps$ is the interaction with the self-induced gauge vector potential.
Treating this interaction as weak, we split $\rho = \rho_0 + \delta \rho$, such that $i d \rho_0 / dt = [H_0,\rho_0]$. Because photoionization
has created electron-hole pairs in coherent superposition of states, the unperturbed density matrix is non-diagonal in a Bloch state basis and takes the form
\begin{equation}
 [\rho_0(t)]_{nn' \ks} = \xi_{nn' \ks} e^{-i \omega_{nn'\ks} t},
\end{equation}
where the diagonal elements $\xi_{nn \ks}=f_{n \ks}=\sum_v |\langle n \ks | \psi_{v \ks} (\infty)\rangle|^2$ are the laser-intensity dependent occupation numbers of single particle states, the off-diagonal elements give the interband coherences and $\omega_{nn'\ks}=\varepsilon_{n\ks} -\varepsilon_{n'\ks}$ are the transition frequencies. To first order in perturbation theory, the equation of motion is
\begin{equation}
 i \frac{d \delta \rho}{dt} = [H_0,\delta \rho ] + [V,\rho_0]
\end{equation}
which reads in components
\begin{equation}
 i \frac{d}{dt} \delta \rho_{n n' \ks} =  \omega_{nn' \ks} \delta \rho_{nn'\ks}(t) +
 \sum_{n''} [V_{n n'' \ks}(t) \xi_{n'' n' \ks} e^{-i \omega_{n''n'\ks} t}  - \xi_{nn'' \ks}  V_{n''n' \ks}(t) e^{-i \omega_{nn''\ks} t}]
\end{equation}
Because interband coherences include rapidly varying phases, such terms more or less tend to average to zero, only diagonal terms with $n''= n$  and $n''=n'$ give dominant contribution, such that the equations of motion simplify to
\begin{equation}
 i \frac{d}{dt} \delta \rho_{nn'\ks} \approx \omega_{nn' \ks} \delta \rho_{nn'\ks}+
 V_{nn' \ks} (t) (f_{n' \ks}-f_{n \ks})
\end{equation}
The Fourier transformation of this result gives
\begin{equation}
 \delta \rho_{nn' \ks}(\omega)=\frac{V_{nn' \ks} (\omega) (f_{n' \ks}-f_{n \ks})}{\omega_{n'n\ks} - \omega}
\end{equation}
and therefore the Cartesian components of the current can be written
\begin{equation}
 J_{\alpha}(\omega)=\sigma_{\alpha \beta}(\omega) E_{\beta}(\omega)
\end{equation}
in terms of conductivity tensor of the photoexcited solid
\begin{equation}
 \sigma_{\alpha \beta}(\omega) = \frac{1}{i \omega} \sum_{nn'} \int \frac{d^3 \ks}{4 \pi^3}
    \frac{ (f_{n\ks}-f_{n'\ks}) (p_{\alpha})_{nn' \ks} (p_{\beta})_{n'n \ks}}
   {\omega_{n'n \ks}-\omega}  - \frac{n_0}{i \omega} \delta_{\alpha \beta}  ,
\end{equation}
here $n_0=32/a_0^3$ is the average bulk density of all electrons. The associated  linear susceptibility tensor $\chi_{\alpha \beta}(\omega)=i \sigma_{\alpha \beta}(\omega)/\omega$ of the laser-excited state is divergent when $\omega \rightarrow 0$. To display this divergence, $\chi$ can be split into interband and intraband contributions \cite{Sipe}
\begin{equation}
\label{chi2}
 \chi_{\alpha \beta} = \frac{1}{\omega^2} A_{\alpha \beta} +\frac{1}{\omega} B_{\alpha \beta} + C_{\alpha \beta}
\end{equation}
where
\begin{equation}
\label{ccof}
 C_{\alpha \beta}=
\sum_{nn'} \int_{{\rm BZ}} \frac{d^3 \ks}{4 \pi^3}  \frac{(f_{n \ks}-f_{n' \ks}) (p_{\alpha})_{nn'\ks} (p_{\beta})_{n'n\ks}}{\omega^2_{n'n\ks} (\omega_{n'n\ks}-\omega)}
\end{equation}
is regular at $\omega=0$, and the coefficients in front of the divergent terms are
\begin{equation}
 B_{\alpha \beta}=
\sum_{nn'} \int_{{\rm BZ}} \frac{d^3 \ks}{4 \pi^3}  \frac{(f_{n \ks}-f_{n' \ks}) (p_{\alpha})_{nn'\ks} (p_{\beta})_{n'n\ks}}{\omega^2_{n'n\ks}},
\end{equation}

\begin{equation}
\label{acof}
 A_{\alpha \beta} = - \sum_n \int_{{\rm BZ}} \frac{d^3 \ks}{4 \pi^3} f_{n \ks} \left[ \frac{1}{m^{\ast}_{n \ks}} \right]_{\alpha \beta}.
\end{equation}
In Eq.(\ref{acof}), $[1/m^{\ast}_{n \ks}]_{\alpha \beta}$ are the Cartesian components of the inverse effective mass tensor of band $n$ with crystal momentum $\ks$
\begin{equation}
 \left[ \frac{1}{m^{\ast}_{n \ks}} \right]_{\alpha \beta}  = \delta_{\alpha \beta} - \sum_{n' \ne n} \frac{(p_{\alpha})_{nn'\ks} (p_{\beta})_{n'n\ks}+(p_{\beta})_{nn'\ks} (p_{\alpha})_{n'n\ks}}{\omega_{n'n \ks}}
\end{equation}
The coefficients $B_{\alpha \beta}=0$ vanish identically only if
occupation numbers of the photoexcited carriers remain invariant under time reversal symmetry [$f_{n \ks}=f_{n-\ks}$].
For intrinsic semiconductor, the coefficients $A_{\alpha \beta}$ vanish trivially, because for filled bands, occupation numbers $f^0_{n \ks}=\{0,1\}$ independently on $\ks$. However $A_{\alpha \beta}$ does not vanish  because in general photoexcited carriers are non-uniformly distributed over the Brillouin zone,  so that $\chi_{\alpha \beta}$ does diverge when $\omega \rightarrow 0$. Introducing a dielectric tensor of the photoexcited solid
\begin{equation}
 \epsilon_{\alpha \beta}(\omega)=\delta_{\alpha \beta}+4 \pi \chi_{\alpha \beta} (\omega),
\end{equation}
the Maxwell's equation $\epsilon_{\alpha \beta}(\omega) A_{\beta}(\omega) = 0$ has a nontrivial solution when $\det  \boldsymbol \epsilon (\omega) = 0$,
which specifies the collective eigenmodes of the polarization.
The dielectric tensor can be split into isotropic part and a traceless part associated with the optical anisotropy of the solid $\epsilon_{\alpha \beta}(\omega) =
 \epsilon(\omega) \delta_{\alpha \beta} + \eta_{\alpha \beta}(\omega)$,  where $\epsilon(\omega)={{\rm tr}}  \boldsymbol \epsilon(\omega)/3$ is the dielectric function.  Assuming that that optical anisotropy is weak $\eta \approx 0$, using Eq.(\ref{chi2}), and neglecting the interband contribution of $B$ coefficients, the dielectric function can be written as a sum of a free carrier Drude term
 and a regular part associated with interband transitions
 \begin{equation}
 \label{epsd}
\epsilon(\omega)=\left(1-\frac{\omega^2_p}{\omega^2} \right)+ \frac{4 \pi}{3} {{\rm tr}} {{\bf C}}(\omega)
 \end{equation}
with definition of plasma oscillation frequency of photoexcited carriers
\begin{equation}
 \omega_p = \left(\frac{4 \pi  n_0}{m^{\ast} } \right)^{1/2}
\end{equation}
and a band-averaged inverse effective mass of an electron-hole pair
\begin{equation}
\label{mass} 
 \frac{1}{m^{\ast}}= \frac{1}{n_0} \sum_n \int_{{\rm BZ}} \frac{d^3 \ks}{4 \pi^3} f_{n \ks} \frac{1}{3} {{\rm tr}} \left[ \frac{1}{m^{\ast}_{n \ks}} \right]
\end{equation}
 The field-free effective mass parameter is infinite, because in an insulating state valence bands are filled and there are no mobile electrons. As the laser intensity increases, mobile electrons and holes are generated and transport charge. Because their occupation number distribution is non-uniform in the crystal momentum space, the effective mass becomes finite and it decreases with the increase of laser intensity.

The imaginary part of the dielectric function is
 \begin{equation}
 {{\rm Im}} \epsilon(\omega) =
 \frac{1}{3\pi} \sum_{nn'} \int_{{\rm BZ}} d^3 \ks  (f_{n\ks}-f_{n'\ks}) |\ds_{nn'\ks}|^2 \delta (\omega - \omega_{n'n\ks}) - \pi \omega_p^2 \delta' (\omega)
 \end{equation}
here $\delta'(x)=-\delta(x)/x$ is the first derivative of the Dirac delta function and $\ds_{nn'}= \ps_{nn'}/i \omega_{nn'}$ is the transition dipole moment between non-degenerate bands. The dielectric function obeys the Thomas-Reiche-Kuhn oscillator strength sum rule
\begin{equation}
 \int d \omega  \omega {{\rm Im}} \epsilon(\omega) = \pi \omega_0^2
\end{equation}
where $\omega_0= (4 \pi n_0)^{1/2}$ is the free-electron plasma frequency. Because of that, switching on the interaction with the electromagnetic field, has an effect of transferring spectral weight from the regular part into the Drude peak near $\omega=0$.
For near-infrared laser irradiation, the absorptive part of the dielectric function ${{\rm Im}} \eps(\omega_L)$ is due to free carrier absorption, i.e. when an electron/hole in an excited band
absorbs a single photon and undergoes an energy-conserving transition into another (unoccupied) higher energy state, the requisite momentum for the transition is provided by the lattice ions.

For increasing concentration of electron-hole pairs, the real part of the dielectric function may vanish and pass through zero, which is accompanied  by a strong free-carrier absorption and attenuation of
the electric field inside the solid. The corresponding resonance plasma frequency $\hat{\omega}_p$ is determined from the condition ${{\rm Re}} \eps(\hat{\omega}_p)=0$, i.e. the frequency at which the Drude term tends to cancel the interband contribution (cf. Eq.\ref{epsd}).  Since the interband term exhibits moderate or relatively weak frequency dependence for near infrared wavelengths, it is convenient to introduce a background dielectric constant $\epsilon_b = 1 +(4 \pi/3) {{\rm tr}} {{\bf C}}(\omega_L)$, such that the plasmon resonance occurs when the laser frequency matches the screened plasma frequency $\omega_L \approx \hat{\omega}_p=\omega_p/\sqrt{\epsilon_b}$.  Thus the fluence dependence of the dielectric function of the pumped solid is parametrized by an effective mass and a background dielectric constant  that incorporates the Pauli exclusion principle via the state filling factors $f_{n\ks}-f_{n'\ks}$.

The normal incidence reflectivity of the photoexcited solid is given by the standard Fresnel formula
\begin{equation}
R=\frac{ (1-n)^2 + \kappa^2}{ (1+n)^2 + \kappa^2}
\end{equation}
where $n$ is the refractive index and $\kappa$ is the extinction coefficient. The optical constants are determined from the relations
\begin{equation} 
\label{opt}
{{\rm Re}} \epsilon = n^2 - \kappa^2, \quad {{\rm Im}} \epsilon =2 n \kappa
\end{equation}
The free carrier absorption coefficient $\alpha=2 \omega\kappa/c$ determines the attenuation of an incident laser irradiation inside the solid and $c$ is the speed of light in vacuum.

\section{Numerical results and discussion}
\begin{figure}[h]
\centering\includegraphics[width=0.7\linewidth]{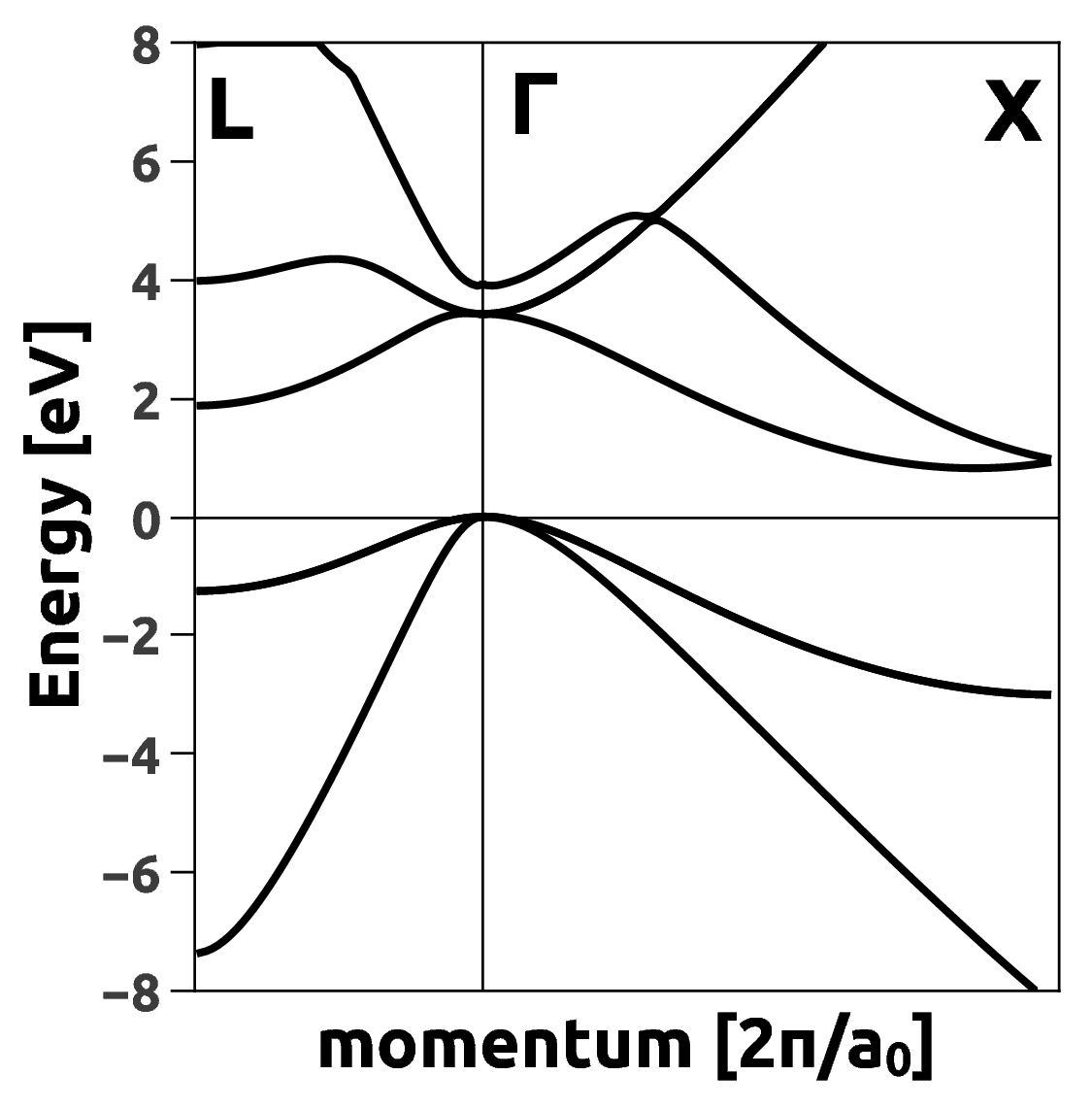}
\caption{Static band structure of silicon along the $\Lambda$ and $\Delta$ lines in the Brillouin zone. The crystal momentum is measured in units $2 \pi/a_0$, where $a_0=$5.43 \AA is the bulk lattice constant.}
\label{fig:Fig1}
\end{figure}

The static energy-band structure of silicon along the $\Delta$ and $\Lambda$ lines in the Brillouin zone is shown in Fig. \ref{fig:Fig1}. The pseudopotential model reproduces quantitatively the principal energy gaps and the optical properties of silicon. The location of the conduction band minimum at $\ks = (0.8, 0, 0)$ relative to the valence
band top at the $\Gamma$ point, specifies the threshold for indirect transitions $\approx$ 1 eV. The threshold for direct transitions is assigned
to the $\Gamma_{25} \rightarrow \Gamma_{15}$ energy gap (3.2 eV).


In the practical calculations \cite{Lagomarsino,Apostolova}, dense sampling of the  Brillouin zone  was made by a Monte Carlo method using 2500 quasi-randomly generated $\ks$-points (points were generated from three-dimensional Sobol sequence in a cube of edge length $4 \pi/a_0$, \cite{Press});  $N$= 20 bands were included in the expansion of the wave-packet over static Bloch orbitals, with 4 valence and 16 conduction bands. The Cayley transformation was applied to propagate the Bloch wave-functions forward in time for small equidistant time steps $\delta t \approx 1$ attosecond
\begin{equation}
 |\psi_{v \ks}(t+\delta t) \rangle = [1-i H(\ks,t) \delta t/2]^{-1} [1+i H(\ks,t) \delta t/2]  |\psi_{v \ks}(t) \rangle
\end{equation}
where $H(\ks,t)=\exp(i \ks \cdot \rs) H(t) \exp(-i \ks \cdot \rs)$, next the single-particle density matrix is evaluated
\begin{equation}
 \rho_{nn' \ks}(t) =\sum_v \langle n \ks| \psi_{v \ks} (t) \rangle
 \langle \psi_{v \ks} (t) | n' \ks \rangle
\end{equation}
and the current density
\begin{equation}
 \jc(t) = {{\rm tr}} [\rho(t) \vs(t)]
\end{equation}
is used to update the macroscopic vector potential according to
\begin{equation}
 \ac_{{\rm ind}}(t+\delta t) = 2 \ac_{{\rm ind}}(t) - \ac_{{\rm ind}}(t-\delta t) + 4 \pi \mathbf{J}(t) \delta t^2
\end{equation}

\begin{figure}[h]
\centering\includegraphics[width=1\linewidth]{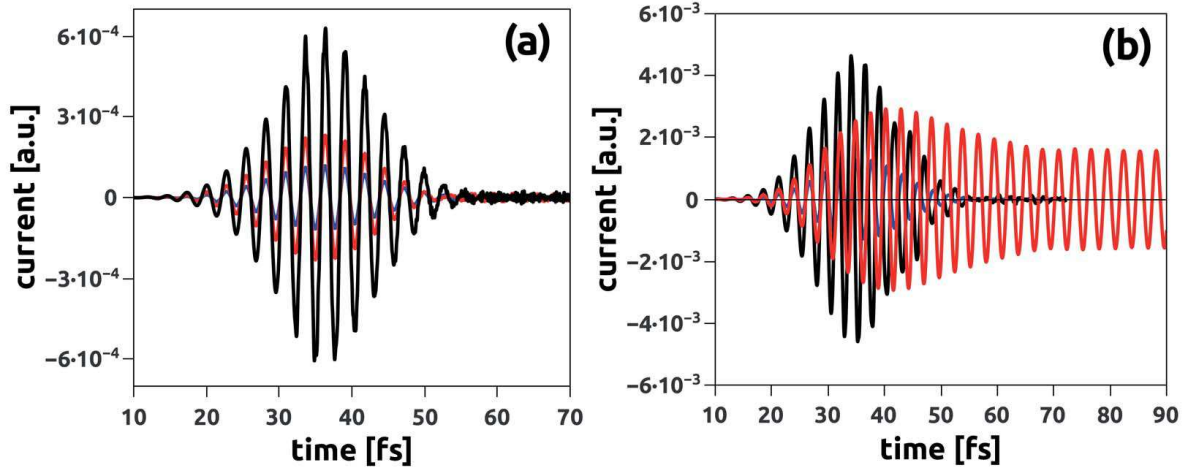}
\caption{{(a) Time evolution of the photoinduced current in bulk silicon subjected to pulsed laser with peak intensity $I=3 \times 10^{13}$ W/cm$^2$ (dotted line), $I=1.2 \times 10^{14}$ W/cm$^2$ (dashed line) and $I=6 \times 10^{14}$ W/cm$^2$ (solid line). (b) Time evolution of the interband current in bulk silicon for laser intensity $I=3 \times 10^{13}$ W/cm$^2$ (dotted line), $I=1.2 \times 10^{14}$ W/cm$^2$ (dashed line) and $I=6 \times 10^{14}$ W/cm$^2$ (solid line). In Fig.(a-b) the pulse is linearly polarized along the [001] direction, the laser wavelength is 800 nm and the pulse duration is 12fs.}
}
\label{fig:Fig2}
\end{figure}

Fig.\ref{fig:Fig2} (a-b) show the laser-induced AC currents in bulk silicon for three different laser intensities $I=3 \times 10^{13}$ W/cm$^2$, $I=1.2 \times 10^{14}$ W/cm$^2$ and $I=6 \times 10^{14}$ W/cm$^2$. The laser wavelength was set to 800 nm (photon energy $\hbar \omega=1.55$ eV) and the polarization vector $\es$ is pointing along the [001] direction. Fig.\ref{fig:Fig2}a gives the component of the total photocurrent in direction of the laser polarization. The induced currents are in phase and follow the Gaussian profile and the periodicity of the applied laser vector potential. The amplitude of the photo-current increases moderately with the increase of the pulse intensity.
Fig.\ref{fig:Fig2}b shows the time evolution and intensity dependence of the intraband part of the photocurrent, the interband part exhibits similar temporal variation (not shown), but is out of phase with the intraband current during the whole time evolution. For the lowest laser intensity shown, $I=3 \times 10^{13}$ W/cm$^2$, completely reversible currents are generated inside the bulk. For higher intensity $I=1.2 \times 10^{14}$ W/cm$^2$, a phase shift of the current relative to the applied laser field gradually accumulates between the peak and the end of the pulse. Intraband motion of charge carriers persists long after the end of the pulse, indicating that irreversible and highly efficient transfer of laser energy to electrons has occurred.  For the highest intensity shown, $I=6 \times 10^{14}$ W/cm$^2$, the photo-current builds up rapidly on the leading edge of the pulse. Shortly after the pulse peak the non-linear phase shift stabilizes and the applied electric field lags behind the current by $\pi/2$.

\begin{figure}[h]
\centering\includegraphics[width=1\linewidth]{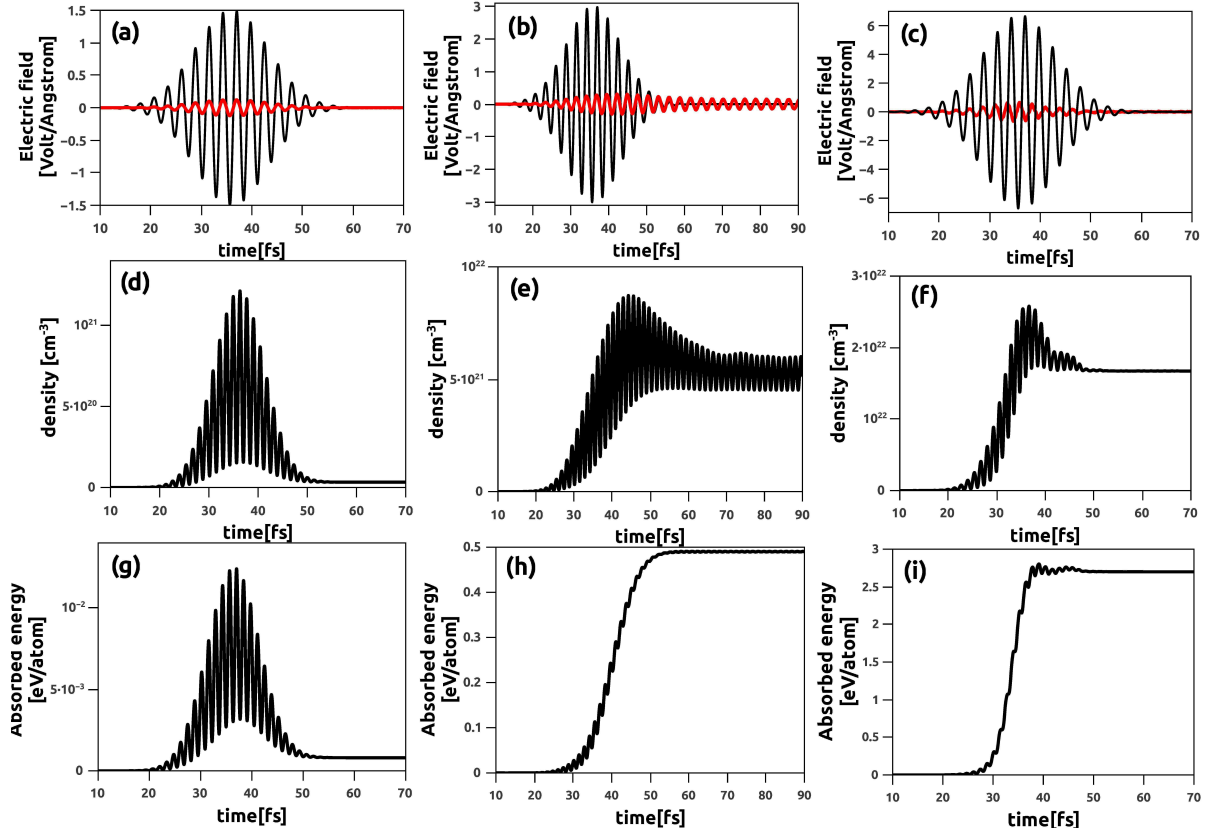}
\caption{{(a-c) Time evolution of the pulsed electric field (in V/\AA) of the applied laser (dashed line) and the total electric field (in V/\AA) (solid line) in bulk silicon. The laser intensity at the pulse peak is 3 $\times $ 10$^{13}$ W/cm$^2$ in (a), 1.2 $\times $ 10$^{14}$ W/cm$^2$ in (b) and is 6 $\times $ 10$^{14}$ W/cm$^2$ in (c). Fig.(d-f) time evolution of the conduction electron density (in cm$^{-3}$) in bulk silicon. Fig.(g-i) give the time evolution of the absorbed energy in eV per Si atom. The laser is linearly polarized along the [001] direction, the laser wavelength is 800 nm and the pulse duration is 12fs.}
}
\label{fig:Fig3}
\end{figure}

Figs. \ref{fig:Fig3}(a-c) show the temporal profile of the applied and total electric fields, corresponding to the three peak laser intensities discussed above. Figs.\ref{fig:Fig3} (d-f) and Figs. \ref{fig:Fig3}(g-i) present the associated time evolution of the conduction electron density and absorbed energy per Si atom, respectively.
In Fig.\ref{fig:Fig3}(a) the total and applied electric fields are in phase during the whole time evolution and dielectric response of electrons is exhibited at low laser intensity: the electric field is screened inside the bulk $\ec=\ec_{{\rm ext}}/\epsilon$ with $\epsilon \approx 12$ - the static dielectric constant of Si. The peak electric field strength inside the bulk reaches 0.1 V/\AA, which is relatively weak to produce high level of electronic excitation. The conduction electron density in Fig.\ref{fig:Fig3}(d) exhibits temporal oscillations following the periodicity of driving laser due to virtual transitions in and out of the conduction band. Real electron-hole are born at the extrema of the laser field by perturbative three- and four-photon photon transitions from the valence into the conduction band. Near the pulse peak the transient number density of electron-hole pairs reaches 10$^{21}$ cm$^{-3}$. This number is subsequently reduced mainly due to disappearance of virtual population. The final photoelectron yield does not exceed 10$^{20}$ cm$^{-3}$. Quite similarly the time evolution of the absorbed energy, shown in Fig.\ref{fig:Fig3}(g), displays rapid transient oscillations due to continuous population and depopulation of the conduction band during each half-cycle. The energy gained during each half-cycle is almost fully returned back to the field, the net energy irreversibly transferred to electrons is well below $0.01$ eV/atom after the end of the pulse .

With increase of laser intensity to $I=1.2 \times 10^{14}$ W/cm$^2$, the applied field progressively lags behind the total electric field after the pulse peak. That is because once a critical number of conduction electrons is generated via multiphoton and tunnel ionization, no band gap exists for conduction electrons and they subsequently undergo resonant transitions to higher lying conduction band states via one photon absorption. 
Few femtoseconds after the peak of the applied laser field, the peak of the total field strength inside the bulk reaches 0.3 V/\AA. This delayed response of the non-linear polarization manifests in self-sustained electric field with strength 0.1 V/\AA ~ long after conclusion of the pulse. In this regime, Fig.\ref{fig:Fig3}(e), photoionization is very efficient: when $5 \%$ of the valence electrons are ionized by multiphoton and tunnel transitions, the collective plasma oscillations of the electron gas emerge. The screened plasma oscillation frequency matches the frequency of the applied laser pulse. The coherent oscillation in the population of the conduction bands is sustained by the non-linear polarization long after the end of the pulse. The electronic excitation energy  increases in cumulative manner following the successive cycles of the driving pulse (cf.  Fig.\ref{fig:Fig3}(h). The laser energy irreversibly deposited to the electronic system rises above the thermal melting threshold of Si (1687 Kelvins) shortly after the pulse peak. A few femtoseconds later, energy deposition saturates and reaches 0.5 eV/atom that exceeds several times the thermal melting threshold of Si.

When the laser intensity is increased further to $I=6 \times 10^{14}$ W/cm$^2$, Fig.\ref{fig:Fig3}(c), the applied field lags behind the total electric field on the leading edge of the pulse, which results in shock-like deposition of energy to electrons, cf.  Fig.\ref{fig:Fig3}(i). During this time interval of 2-3 laser cycles, the conduction electron density in increases rapidly by more than two orders of magnitude, Fig.\ref{fig:Fig3}(f). Prior to the pulse peak, the total and applied electric fields are completely out of phase, and further energy transfer to electrons terminates. The number of conduction electrons and the excitation energy saturate to 2 $\times $ 10$^{22}$ cm$^{-3}$ and 3 eV/atom, respectively. For such high level of excitation, metallic response of the photoexcited Si is clearly exhibited: conduction electrons screen out the electric field inside the bulk. 

\begin{figure}[h]
\centering\includegraphics[width=1\linewidth]{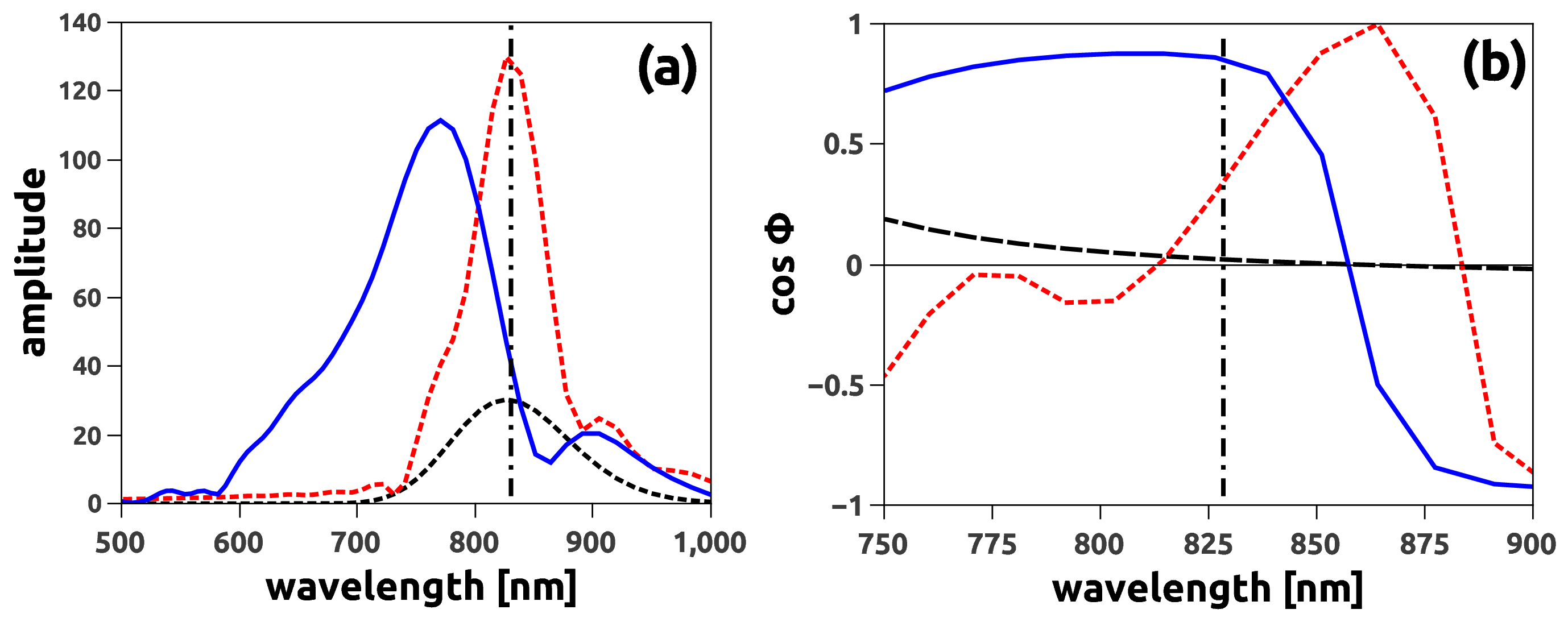}
\caption{{ Fig.(a) Laser intensity dependence of the spectral amplitude and (b) (cosine of) the spectral phase of the pulsed electric field inside bulk silicon.  Different curves correspond to different laser intensities: $I=3 \times 10^{13}$ W/cm$^2$ (dashed line), $I=1.2 \times 10^{14}$ W/cm$^2$ (dotted line) and $I=6 \times 10^{14}$ W/cm$^2$ (solid line). In Fig.(a-b) the laser is linearly polarized along the [001] direction, the laser wavelength is 800 nm and the pulse duration is 12fs. }
}
\label{fig:Fig4}
\end{figure}

To analyze this further, in Fig.\ref{fig:Fig4}(a) we plot the Fourier transform of the pulsed electric field. In the low laser intensity regime, the distribution is centered at the laser wavelength ($\approx$ 800 nm) and exhibits Gaussian profile. The spectral distribution gains large amplitude and becomes narrower when the peak laser intensity is increased to the region of the plasmon resonance with $I \approx$ 10$^{14}$ W/cm$^2$. For the highest intensity shown, 6 $\times$ 10$^{13}$ W/cm$^2$, the distribution broadens in wavelength resulting in contraction of the pulse length in the time domain. The central wavelength undergoes a progressive blue shift with increasing laser intensity, such that the plasma frequency moves off resonance with the applied laser. The spectral phase of the pulsed electric field relative to the photo-current is also shown in Fig.\ref{fig:Fig4}(b). For the low intensity, the spectral phase shift is close to $\pi/2$ in the absorption region and dielectric response is exhibited. For increased laser intensity in the region of the plasmon resonance, the spectral components on the red side of the central wavelength are in phase with the photocurrent, such that electrons gain much energy by interacting with the laser. Due to the dispersion of the phase in the absorption region, this contribution is partially cancelled by interaction with shorter wavelengths on the blue side of the central wavelength. The rapid variation of the spectral phase also causes a time delay of the envelope of the pulsed electric field relative to applied laser (cf. also Fig.\ref{fig:Fig3}(b). At the highest intensity shown, the phase stabilizes in the absorption region: the pulsed electric field and the current are almost in phase, such that free-carrier absorption and energy transfer to electrons is greatly enhanced.

\begin{figure}[h]
\centering\includegraphics[width=1\linewidth]{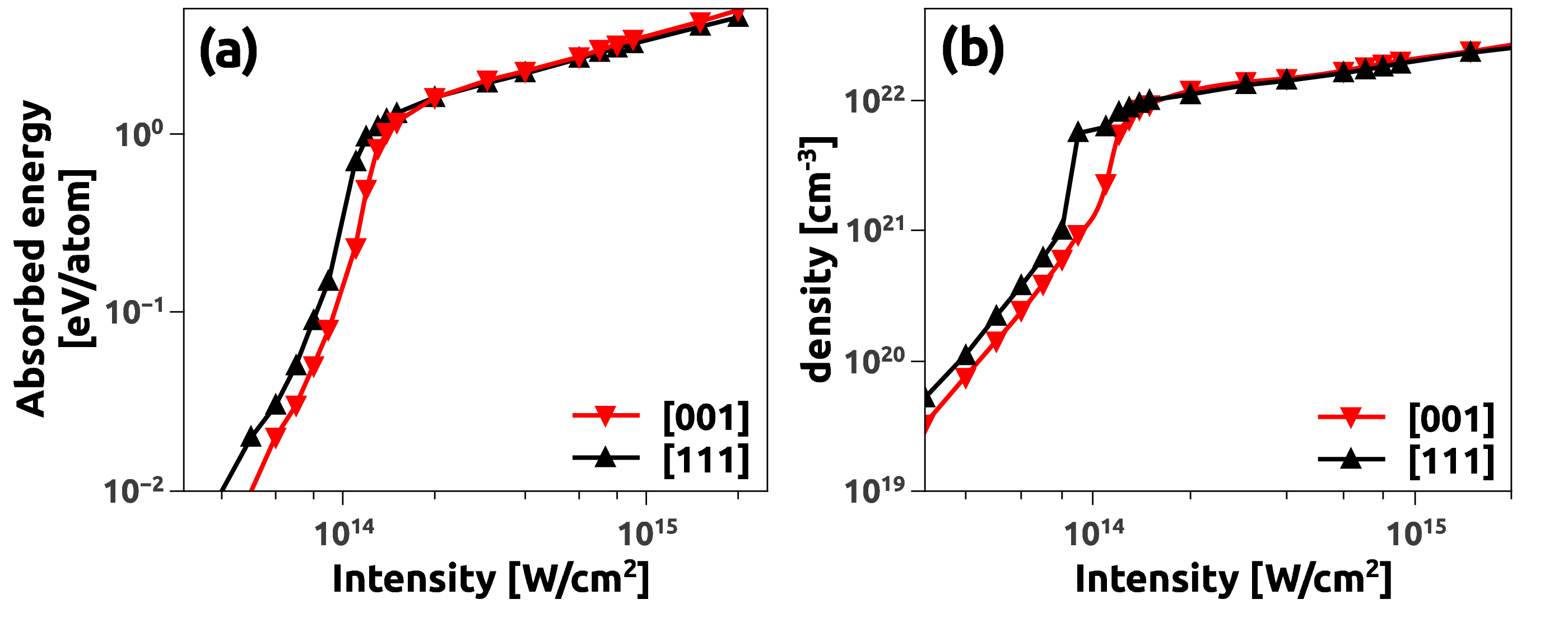}
\caption{{Peak laser intensity dependence of the absorbed  energy per atom (in eV) (a) and (b) the  conduction electron density (in cm$^{-3}$) in bulk silicon subjected to 12fs pulsed laser irradiation with wavelength 800 nm. Upper triangles - laser is linearly polarized along the [111] direction, lower triangles - laser linearly polarized along [001] direction. }
}
\label{fig:Fig5}
\end{figure}

In Fig.\ref{fig:Fig5}(a-b), we plot the dependence of the absorbed energy and conduction electron density on the peak laser intensity for two different laser polarization directions - [001] and [111] directions. For laser intensities $I < 10^{14}$ W/cm$^2$, perturbative superlinear scaling trend is exhibited with $\Delta E \sim I^3$ corresponding to dominant three-photon absorption. When $I > 10^{14}$ W/cm$^2$, a qualitative change in electron dynamics occurs with sublinear scaling trend $\Delta E \sim I^{1/2}$ and $n \sim I^{1/3}$. This regime corresponds to high level of electronic excitation with more than 5 $\%$ of valence electrons promoted into anti-bonding orbitals and absorbed energy $\Delta E > 0.1$ eV/atom. Similar sublinear scaling trend of the photocarrier density in silicon in regime of high excitation was measured and reported in Ref.\cite{Ionin}.  The absorbed energy depends also sensitively on the laser polarization direction: in the low intensity regime, energy is more efficiently deposited for a laser pulse linearly polarized along the [111] direction as compared to the [001] direction. The polarization dependence is most pronounced in the region of the plasmon resonance near $I \approx 10^{14}$ W/cm$^2$, e.g. 0.7 eV/atom are deposited for laser polarized along the [111] direction, while 0.2 eV/atom are deposited when the polarization vector points along the [001] direction. It is worth noting that anisotropy and polarization dependence of multiphoton charge carrier generation rate in diamond has been reported recently using photoluminescence and nonlinear absorption measurements \cite{Kozak}.

\begin{figure}[h]
\centering\includegraphics[width=1\linewidth]{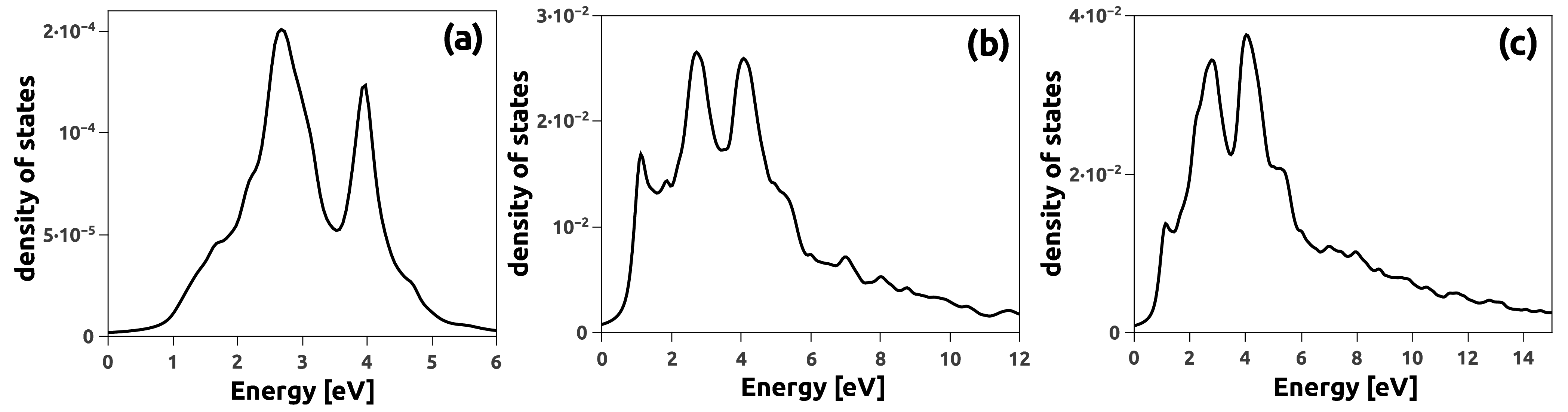}
\caption{{Density of conduction electron states in silicon after excitation by 12fs near-infrared laser with wavelength 800 nm. The laser intensity is: (a) $I=3 \times 10^{13}$ W/cm$^2$, (b) $I=1.2 \times 10^{14}$ W/cm$^2$ and (c) $I=6 \times 10^{14}$ W/cm$^2$. The laser is linearly polarized along the [001] direction.}}
\label{fig:Fig6}
\end{figure}

Further details on the excitation mechanism are included in the energy distribution of photoexcited electrons, Fig.\ref{fig:Fig6}(a-c). The density of conduction band states is characterized by two broad peaks with energies 2.5 and 4 eV above the valence band maximum, which are associated with 3 and 4-photon transitions. The intensity of the distribution increases with the increase of the laser intensity and high energy tail emerges. In this high intensity regime, when photoionization saturates due to altered probabilities for interband transitions, some fraction of electrons from the low energy part of the distribution spill out into the high energy tail as a result of free carrier absorption. This redistribution of conduction electrons over states is regulated by the Pauli exclusion principle.

\begin{figure}[h]
\centering\includegraphics[width=1\linewidth]{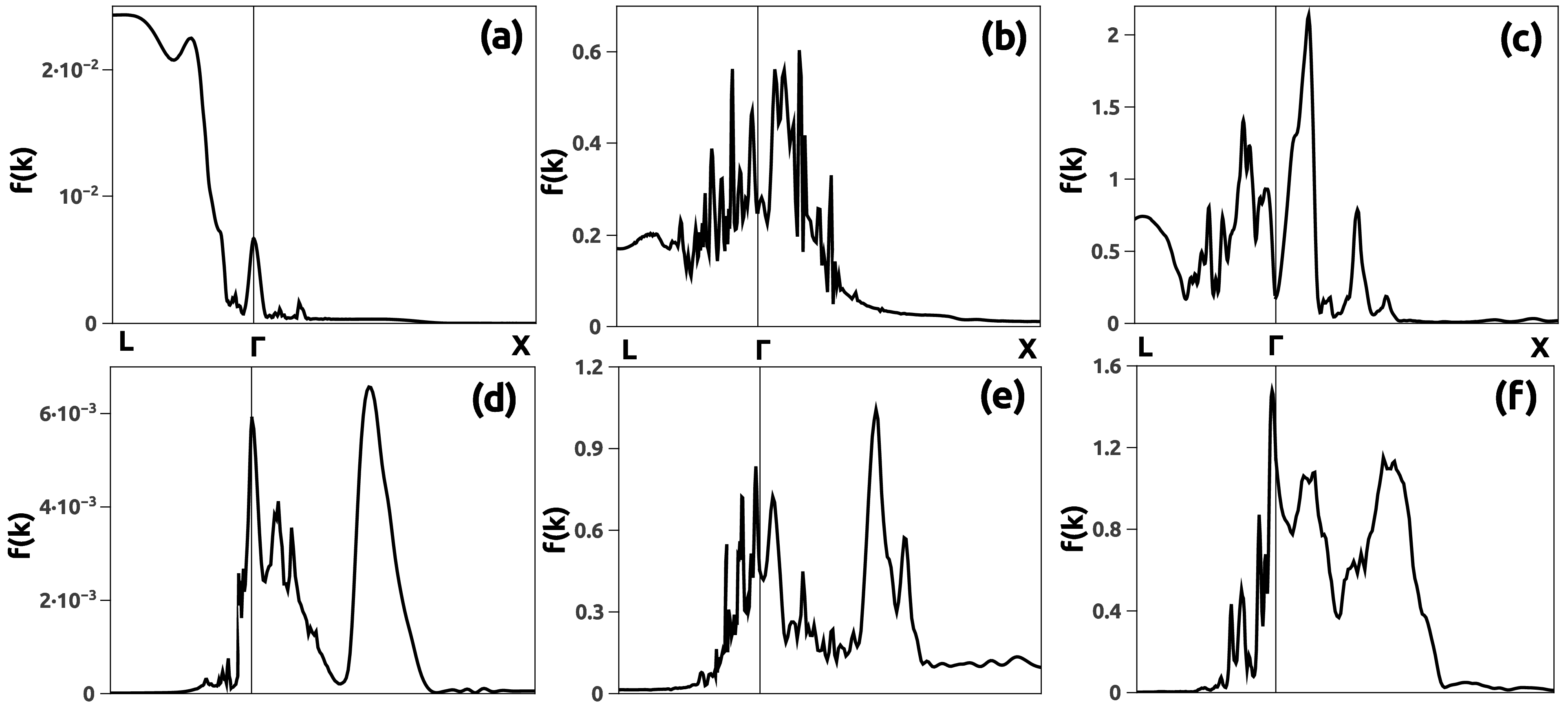}
\caption{{Crystal momentum distribution of conduction electrons in silicon after excitation by 12fs near-infrared laser pulse with wavelength 800 nm. In Fig.(a-c) the laser is linearly polarized along the [001] direction. In Fig.(d-f) the laser polarization vector points along the [111] direction. The peak laser intensity is: $I=3 \times 10^{13}$ W/cm$^2$ in Fig.(a) and Fig.(d), $I=1.2 \times 10^{14}$ W/cm$^2$ in Fig.(b) and Fig.(e), and $I=6 \times 10^{14}$ W/cm$^2$ in Fig.(c) and Fig.(f).}
}
\label{fig:Fig7}
\end{figure}

The momentum-resolved distribution function of conduction electrons $f(\ks)=\sum_c f_c(\mathbf{\ks})$ is shown in Fig.\ref{fig:Fig7} for the three different laser intensities. Figs.\ref{fig:Fig7}(a-c) correspond to laser linearly polarized along the [001] direction and Figs.\ref{fig:Fig7}(d-f) refer to the [111] direction. 
The electron distribution is highly non-uniform and anisotropic in the Brillouin zone. For low intensity $I=3 \times 10^{13}$ W/cm$^2$, small fraction of valence electrons are promoted into the conduction bands with occupation numbers of order $10^{-3}$ (cf. Fig.\ref{fig:Fig7}(a,d). More electrons emerge along the $\Lambda$ line for laser polarized along [001] direction, Fig.\ref{fig:Fig7}a. When the laser polarization vector points in the [111] direction, Fig.\ref{fig:Fig7}d, spectral structure is exhibited near the Brillouin zone center and a broad shoulder of electrons emerges along the $\Delta$ line and near the X valleys. As the laser intensity increases to $I=1.2 \times 10^{14}$ W/cm$^2$, the distribution broadens in the Brillouin zone and the population of conduction bands increases substantially with occupation numbers $f_{c k} \sim$ 1,  cf. Fig.\ref{fig:Fig7}(b,e).  For the highest intensity shown $I=6 \times 10^{14}$ W/cm$^2$, Fig.\ref{fig:Fig7}(c,f), the occupation of conduction bands does not change much on average, instead electrons redistribute over energy levels obeying the Pauli's exclusion principle.

\begin{figure}[h]
\centering\includegraphics[width=1\linewidth]{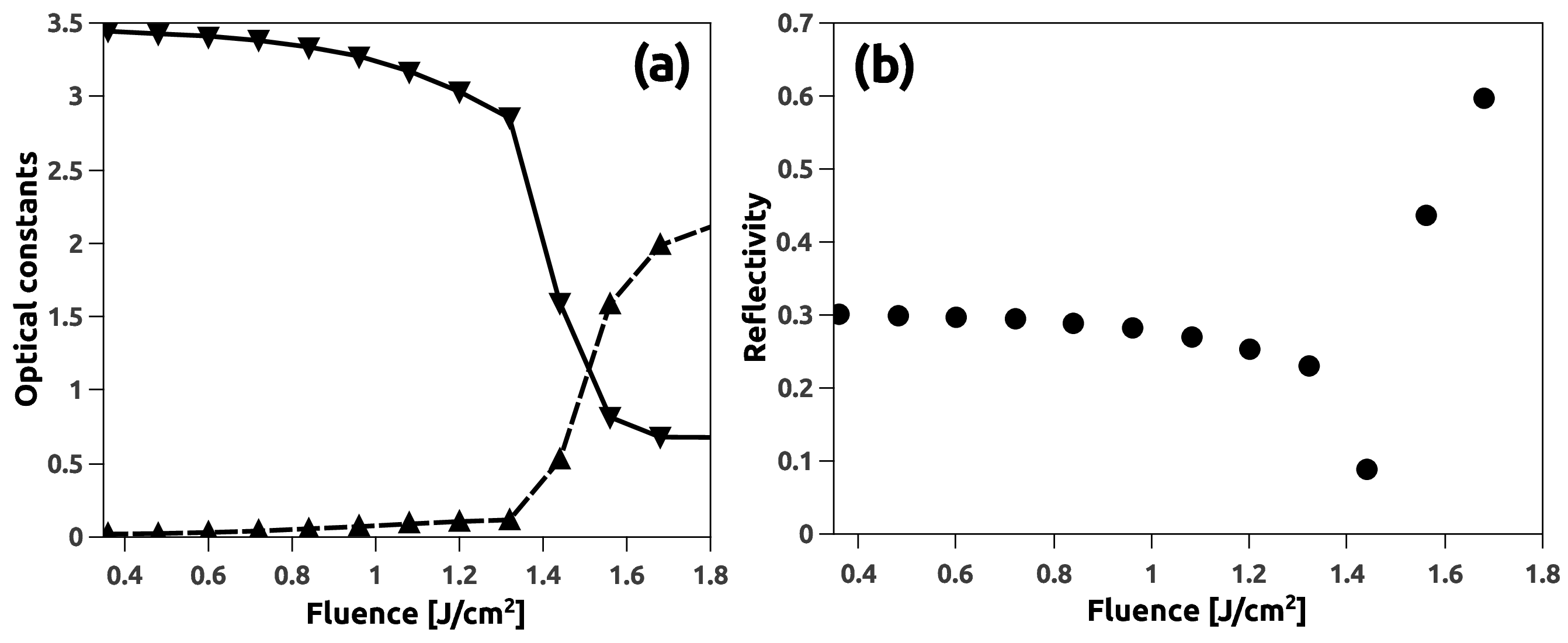}
\caption{{ (a) Laser fluence dependence of the refractive index (lower pointing triangles) and the extinction coefficient (upper pointing triangles) of photoexicted silicon and Fig. (b) gives the change of normal incidence reflectivity as a function of the fluence. In Fig.(a-b) the laser is linearly polarized along the [001] direction, the laser wavelength is 800 nm and the pulse duration is 12fs.}
}
\label{fig:Fig8}
\end{figure}

Using these distributions we compute the optical constants of the photoexcited Si. The change of the refractive index, the extinction coefficient and the normal incidence reflectivity as a function of the laser fluence are shown in Fig.\ref{fig:Fig8}(a-b) for laser linearly polarized along the [001] direction. For low fluences, below 0.4 J/cm$^2$, the refractive index $n=3.5$ and the reflectivity $R=0.3$ are nearly constants and ground state dielectric response is reproduced. The refractive index and the reflectivity decrease with the increase of the fluence, while the extinction coefficient increases.
The screened plasma frequency $\hat{\omega}_p$ appears near the minimum of the reflectivity curve at fluence 1.4 J/cm$^2$, when $n \approx \kappa$. Thus the refractive index is small near $\hat{\omega}_p$ and the reflectivity is also small. Above that fluence, the reflectivity increases rapidly and reaches $R=0.6$, the steepness of the slope in the reflectivity rise above that fluence is governed by the extinction coefficient and free-carrier absorption.  This behavior reproduces qualitatively the main features of the measured evolution of reflectivity with fluence \cite{Sokol} and is a result of competition between band filling, Pauli blocking and free carrier response: for fluences close to the threshold fluence 1.4 J/cm$^2$, the interband polarization tends to increase relative to the ground state due to softening of Si-Si bonds, while since free-carrier polarization gives always negative contribution it reduces the high polarizability of the pumped solid. When the screened plasma frequency matches the laser frequency, the free-carrier response dominates, the dielectric constant vanishes and electron-hole pairs exhibit a coherent plasma oscillation. For fluences above the threshold fluence, free-carrier absorption dominates the optical properties (with $\kappa > n$), the photoexcited crystal becomes dissipative and strongly attenuates the subsequent incident irradiation inside the solid. For instance the  penetration depth (the thickness of the plasma layer at the surface) $\delta = \lambda/(2 \pi \kappa) \approx$ 30 nm becomes much smaller than the laser wavelength. 

Regarding the optical constants of photoexcited Si, it may be worth to comment on the numerical pump-probe data for silicon irradiated with near-infarared wavelengths presented in Ref.\cite{Sato}. The calculated dielectric response of photoexcited Si  was compared with a free-carrier Drude model. The real part of the dielectric function is well fitted by the model, when treating the effective mass as a fitting parameter and taking the conudction electron density from numerics. The fitted effective mass was found to increase with increasing pump field intensity, i.e. charge carriers should become less mobile and thus respond less sensitively to strong electric fields. In contrast our calculation predicts that the effective mass 
decreases monotnocially with the increase of the pump fluence and charge carriers tend to increase their mobility. This difference steems from due Drude weight, which specifies the DC  limit ($\omega \rightarrow 0$) of the conductivity 
$\sigma(\omega) \rightarrow D \delta(\omega)$ (cf. also Eq. (\ref{epsd})). The Drude weight $D = \pi n_0/m^{\ast}$ is determined by the ratio of the all electron density $n_0$ to the effective mass of photoexcited carriers. Instead the Drude weight according to Ref.\cite{Sato} is $\pi n_{eh}/m^{\ast}$, where $n_{eh}$ is the density of photoexcited electron-hole pairs which increases with the increase of the pump fluence and since $m^{\ast}$ is treated as a fitting parameter it can take any value. For $I=1.2 \times 10^{14}$ W/cm$^2$,  
we obtain $m^{\ast} \approx 13 m_e \gg m_e$, while the reported effective masses in Ref.\cite{Sokol,Sato} are in general smaller than the free-electron mass. However since $n_{eh}$ in Si usually much smaller than $n_0=2 \times 10^{23}$ cm$^{-3}$, it explains the reported small values for the effective masses in photoexcited Si obtained from fits to Drude model.  

\begin{figure}[h]
\centering\includegraphics[width=1\linewidth]{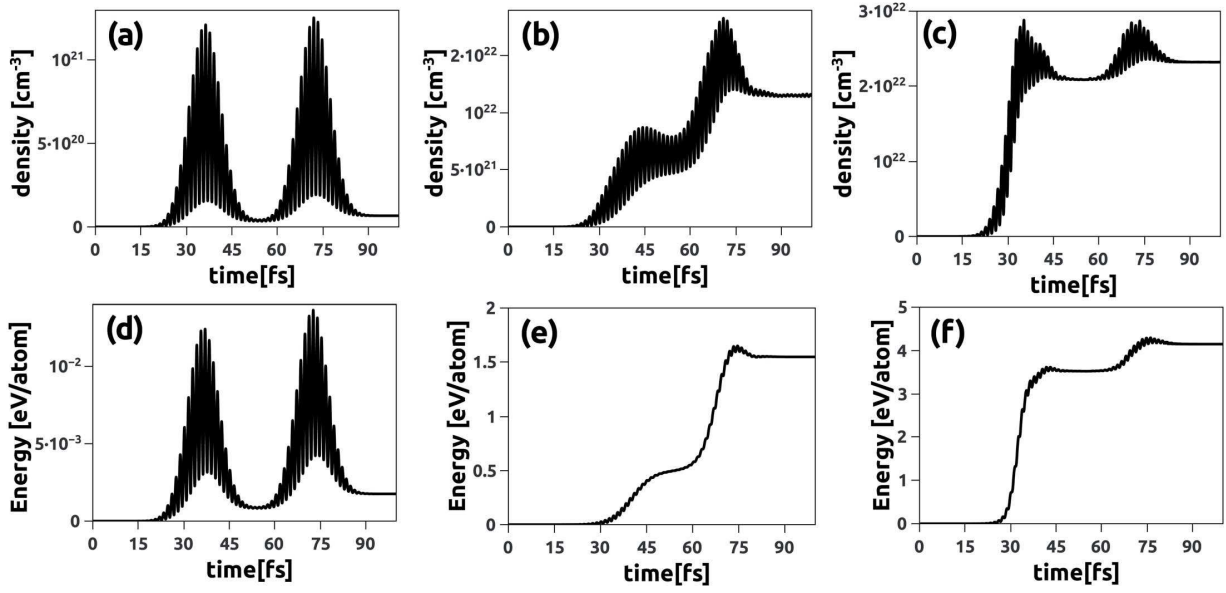}
\caption{{ (a-c) Time-evolution of the conduction electron density (in cm$^{-3}$)  and (d-f) absorbed energy per atom (in eV) in bulk silicon subjected to a sequence of two identical non-overlapping 12fs near-infrared pulses with wavelength 800 nm. The peak laser intensity $I=3 \times 10^{13}$ W/cm$^2$ in Fig.(a) and Fig.(d), $I=1.2 \times 10^{14}$ W/cm$^2$ in Fig.(b) and Fig.(e), and $I=6 \times 10^{14}$ W/cm$^2$ in Fig.(c) and Fig.(f). }
}
\label{fig:Fig9}
\end{figure}

To investigate further the nonlinear response of the photoexcited plasma to strong laser field we apply a second identical pulse after the excitation of Si by the first pulse. While for low intensities, Fig.\ref{fig:Fig9}(a,c), the effect of two pulses is additive in terms of absorbed energy and photoelectron density, strong effect of enhancement in terms of absorbed energy occurs for laser intensity in the region of the plasmon resonance in Fig.\ref{fig:Fig9}(b,d): the second pulse deposits two times more energy than the first one.  On the leading edge of the second pulse, wave functions of valence electrons include admixtures from conduction bands, which enables 
resonant one-photon transitions into higher-energy conduction bands, thereby increasing the photoelectron yield; the Pauli exclusion principle regulates the energy exchange between the light field and the electrons.
For increased pulse intensity, Fig.\ref{fig:Fig9}(c,e) the plasma oscillation frequency shifts above the laser frequency, the laser-excited state displays metal-like response and is reflective for the incident irradiation and a large portion of its energy is reflected away from the silicon surface.

\section{Conclusion}

In summary we have investigated the ultrafast photoexcitation and plasma formation in crystalline silicon interacting with intense 12fs near infrared laser pulse. For peak laser intensities near and above $10^{14}$ W/cm$^2$, photoionization creates dense plasma of electron-hole pairs and strongly absorbing state of silicon for near infrared laser wavelengths, with deposited energies a few times exceeding the thermal melting threshold. This state may be considered as a precursor to ultrafast phase transition and melting of silicon, as reported experimentally using femtosecond lasers. The fluence dependence of the normal incidence reflectivity of the pumped silicon reproduces qualitatively the principal features observed in the experiments. We have demonstrated the relative importance of band filling, Pauli blocking and Drude free-carrier responses.
The optical anisotropy of the photoexcited silicon was considered as weak, effects like optically induced birefringence may be subject to our follow-up paper.

\section*{Acknowledgements}
This material is based upon work supported by the Air
Force Office of Scientific Research under award number FA9550-19-1-7003.Cost action CA17126 is acknowledged.
(B.O.) also  acknowledges financial support from the Bulgarian
National Science Fund under Contract No. 08-17.

\end{document}